\documentclass[journal]{IEEEtran}
\IEEEoverridecommandlockouts

\usepackage{cite}
\usepackage{amsmath,amssymb,amsfonts}
\usepackage{algorithmic}
\usepackage{graphicx}
\usepackage{textcomp}
\usepackage{xcolor}
\usepackage{tabularx}
\usepackage{diagbox}
\usepackage{siunitx}
\usepackage{adjustbox}
\usepackage{siunitx}
\usepackage{multirow}
\usepackage{booktabs}
\usepackage{array}
\usepackage{multirow}
\usepackage{setspace} 
\usepackage{arydshln}

\begin{document}

\title{Knowledge-Based Deep Learning for Time Efficient Inverse Dynamics
\\
\author{Shuhao Ma,
        Yu Cao,
        Ian D. Robertson,~\IEEEmembership{Fellow, IEEE},
        Chaoyang Shi,~\IEEEmembership{Member, IEEE},
        Jindong Liu,
        Zhi-Qiang Zhang,~\IEEEmembership{Member, IEEE}}

\thanks{This work was supported in part by the China Scholarship Council (CSC) under Grant 202208320117. (Corresponding author: Zhi-Qiang Zhang)}
\thanks{Shuhao Ma, Yu Cao, Ian D. Robertson, and Zhi-Qiang Zhang are with the School of Electronic and Electrical Engineering, University of Leeds, Leeds LS2 9JT, U.K. (e-mail: elsma@leeds.ac.uk; y.cao1@leeds.ac.uk; i.d.robertson@leeds.ac.uk; z.zhang3@leeds.ac.uk).}
\thanks{Chaoyang Shi is with the Key Laboratory of Mechanism Theory and Equipment Design of Ministry of Education, School of Mechanical Engineering, Tianjin University, Tianjin 300072, China, and also with the International Institute for Innovative Design and Intelligent Manufacturing of Tianjin University in Zhejiang, Shaoxing 312000, China (e-mail: chaoyang.shi@tju.edu.cn).}
\thanks{Jindong Liu is with the Estun Medical Ltd, Nanjing 211102, China (e-mail:liujindong@estun.com)}}

\maketitle

\begin{abstract}
Accurate understanding of muscle activation and muscle forces plays an essential role in neuro-rehabilitation and musculoskeletal disorder treatments. 
Computational musculoskeletal modeling has been widely used as a powerful non-invasive tool to estimate them through inverse dynamics using static optimization, but the inherent computational complexity results in time-consuming analysis.
In this paper, we propose a knowledge-based deep learning framework for time-efficient inverse dynamic analysis, which can predict muscle activation and muscle forces from joint kinematic data directly while not requiring any label information during model training.
The Bidirectional Gated Recurrent Unit (BiGRU) neural network is selected as the backbone of our model due to its proficient handling of time-series data. Prior physical knowledge from forward dynamics and pre-selected inverse dynamics based physiological criteria are integrated into the loss function to guide the training of neural networks.
Experimental validations on two datasets, including one benchmark upper limb movement dataset and one self-collected lower limb movement dataset from six healthy subjects, are performed. The experimental results have shown that the selected BiGRU architecture outperforms other neural network models when trained using our specifically designed loss function, which illustrates the effectiveness and robustness of the proposed framework.
\end{abstract}


\IEEEpeerreviewmaketitle

\section{Introduction}\label{intro}
Knowledge of the underlying interactions between neuromuscular and skeletal systems is crucial for many applications, ranging from exoskeleton and prosthetics development \cite{9923768, 8627981}, quantitative biomechanical evaluation \cite{9116986}, to musculoskeletal (MSK) disease treatments \cite{10388411} and sports performance optimization\cite{9546647, 10050735}.
It is challenging to measure key variables such as muscle forces and muscle activation in vivo directly\cite{10538305,9756369, 10366325}, thus 
computational MSK modeling, which employs forward and inverse dynamics, has emerged as a powerful practical alternative solution to estimate these variables\cite{NeuromusculoskeletalModel}. Forward dynamics is an effective method for mapping neural command which can be measured by electromyogram (EMG) sensors to muscle forces and joint motion, via muscle activation dynamics, muscle contraction dynamics, and skeletal dynamics \cite{NeuromusculoskeletalModel, 9258965,8681204}.
However, the precise measurement of EMG signals is not always feasible, particularly for the muscles located deep within the human body \cite{10.1093/ptj/61.1.7, 8681204}. 

Unlike forward dynamics, inverse dynamics address this issue from the opposite side, which utilizes readily available joint kinematics and external forces as inputs to calculate joint torque and muscle forces \cite{NeuromusculoskeletalModel, 9672677, 10197487}.  However,  there are countless number of neural commands that can lead to the same movement. To solve this redundancy problem, static optimization (SO) has been widely applied so far \cite{SEIREG197589}.
The main principle of SO is to formulate the problem as an optimization task, with the objective of finding muscle forces that satisfy task constraints (e.g., joint motion) while minimizing pre-selected physiological criteria in each single time step \cite{CROWNINSHIELD1981793}. 
For instance, Heintz et al. used SO to estimate individual muscle forces during gait via minimizing muscle stress in a single gait cycle and compared the outcomes with the EMG-to-force processing method \cite{HEINTZ2007279}. Zargham et al. accurately predicted muscle activities and knee contact forces during walking by minimizing muscle activation with exponents of three or four through SO \cite{ZARGHAM2019223}.
Veerkamp et al. proposed a new combined performance criterion, integrating the cost of transport, muscle activity, and head stability, to align simulations more closely with experimental data through SO \cite{VEERKAMP2021110530}.
However, the SO method would require recalculating each new set of movements, which can be time-consuming leading to high running latency \cite{TRINLER201955}. 

Recently, with the advancement of artificial intelligence technology, data-driven approaches are emerging as a viable alternative.
In MSK forward modeling, deep learning is extensively applied to map neural commands to desired motion, especially in continuous motion prediction and gesture classification \cite{9927168,10138591,9970372}.
For instance, Fan et al. developed a deep forest-based neural decoder that utilizes limited single-finger surface electromyogram (sEMG) data to accurately predict multi-finger force combinations, demonstrating the potential for efficient and precise neural decoder training in advanced robotic hand control \cite{10398450}.
Zhang et al. introduced the Long Short-Term Memory with Dual-Stage Attention (LSTM-MSA) model to process EMG signals and predict intended actions \cite{10332235}.
In MSK inverse modeling, deep learning is increasingly employed to infer neural commands and muscle forces from observed motions. For instance, Nasr et al. proposed InverseMuscleNET, a machine learning model designed to estimate the pattern of muscle activation signals \cite{10.3389/fncom.2021.759489}. 
Liang et al. presented a novel deep learning model to estimate the EMG envelope during gait using inertial measurement units\cite{LIANG2024112093}.
Dao et al. developed a recurrent deep neural network that incorporates dynamic temporal relationships to estimate muscle forces from kinematics data during a gait cycle \cite{RN118}.
Although these methods facilitate fast estimations essential for applications that require immediate feedback \cite{RN8}, the effectiveness of such models is highly dependent on the quality and representativeness of the labeled training dataset \cite{RN9, 10.3389/fbioe.2020.603907}. 
Obtaining high-quality labeled data for model training would normally involve SO, which is often time-consuming and resource-intensive.
Additionally, most of these models neglect the physical significance of the underlying neuromechanical processes \cite{RN97}.

To address the limitations of data-driven methods and traditional inverse dynamics optimization, this paper introduces a novel knowledge-based deep learning method to map the relationship between joint kinematics, muscle forces, and muscle activation levels. The main contribution of the paper includes:
1) A time-efficient deep learning method based on the Bidirectional Gated Recurrent Unit (BiGRU) for the estimation of muscle activation and forces from joint kinematic data directly is developed.
2) Prior physical knowledge from forward dynamics and pre-selected inverse dynamics based physiological criteria are integrated into the loss function to guide the training of neural networks, and
3) No label information is required during the deep neural network training.
To validate the proposed method, a self-collected dataset involving knee bending and a benchmark dataset comprising elbow flexion/extension (FE) are used in the experiments. 
Our model can achieve average $R^2$ values of $0.945$ for knee bending and $0.941$ for elbow FE, thereby validating the effectiveness and generalizability of our proposed method.
Additionally, utilizing the same BiGRU backbone architecture, our proposed method demonstrates comparable performance to traditional supervised data-driven methods, further validating the effectiveness of our proposed method.

The remainder of this paper is organized as follows: The configuration of our proposed method and the design of loss functions are introduced in Section \ref{method}. Dataset and experimental preparation are then described in Section \ref{Dataset}. Section \ref{results} reports the experimental results of two datasets, and discussions about the limitations and future research avenues are presented in Section \ref{discussions}. Section \ref{conclusion}, concludes this article.

\begin{figure*}
  \centering
  \includegraphics[scale = 0.6]{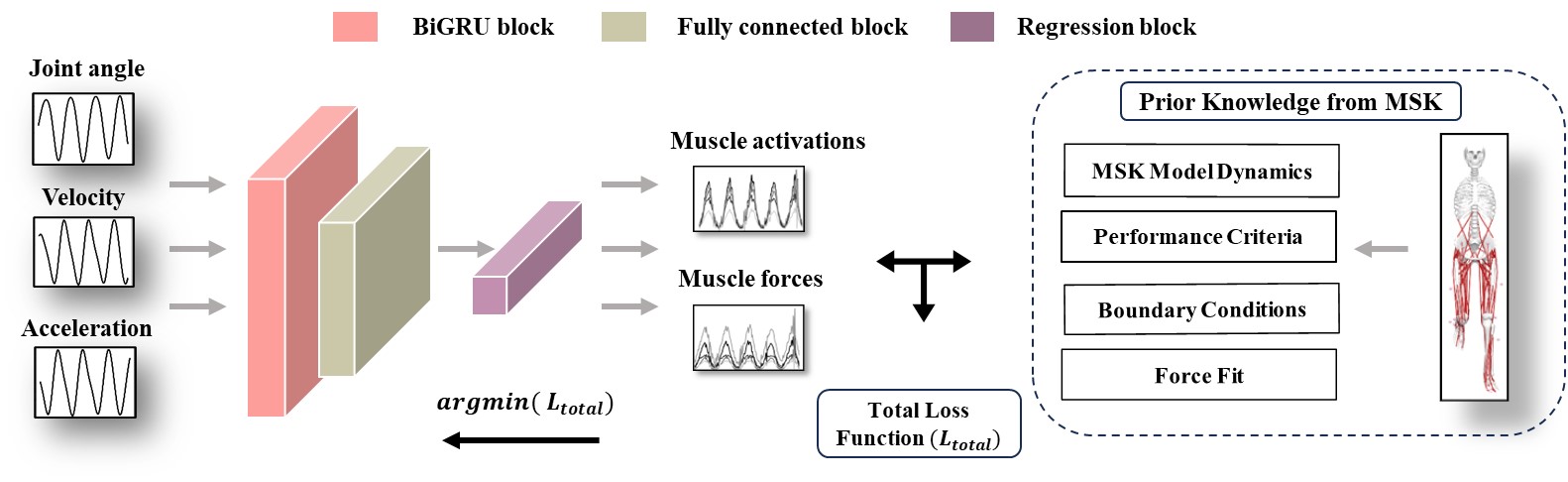} 
  \caption{The main framework of the proposed knowledge-based deep learning framework. Inputs to the neural network comprise joint kinematic data (angles, angular velocities, and angular accelerations), while outputs are muscle activations and forces for the modeled muscles. The total loss, incorporating prior knowledge from MSK modeling, guides the network training to ensure physiologically consistent predictions.}
  \label{mainframe}
\end{figure*}
\section{Methods}\label{method}
In this section, we describe the details of the proposed method, in the context of the estimation of the muscle activations and muscle forces from joint kinematics directly, including the main framework, neural network architecture and training, and the MSK dynamics inspired loss functions. 

\subsection{Main Framework}
Fig. \ref{mainframe} illustrates the primary architecture of our proposed method, which focuses on predicting muscle activation and muscle forces from joint kinematic data directly.
Specifically, in the neural network surrogate, the inputs to the deep neural network comprise joint kinematic data, including the joint motion $q_t$, joint angular velocity $\dot{q_t}$, and joint angular acceleration $\ddot{q_t}$, at the time step $t$.
The outputs comprise the muscle activations $\mathbf{\hat{a}}_t = (\hat{a}_{t,1}, \hat{a}_{t,2},\cdots, \hat{a}_{t, N})$ and muscle forces $\mathbf{\hat{F}}_t = (\hat{F}_{t,1}, \hat{F}_{t,2},\cdots, \hat{F}_{t, N})$, where $N$ represents the total number of muscles at the joint of interest. 
A BiGRU based neural network is employed as the backbone to extract more discriminative features and establish the relationship between the inputs and outputs.
Unlike conventional data-based loss functions, our innovative loss function, inspired by MSK dynamics, is utilized to guide model regression without the reliance on label information.

\subsection{Neural Network Architecture and Training}

As depicted in Fig. \ref{mainframe}, our neural network architecture consists of three main components: a BiGRU block for temporal feature extraction, a fully connected block for feature transformation, and a regression block for final prediction.
The BiGRU block employs a two-layer Bidirectional GRU ($64$ neurons per layer) with a dropout rate of $0.3$ for regularization.
The output of the BiGRU is utilized and fed into the subsequent fully connected block, which transforms these temporal features through two linear layers ($128$ nodes each) with ReLU activation and dropout regularization. 
The regression block then maps the transformed features to muscle activations and forces predictions through a linear layer.
The model is trained using the Adam optimizer with an initial learning rate of $5e-4$, batch size of $64$, and maximum iterations of $2000$, minimizing the proposed knowledge-based loss functions detailed in Section II-C.

\subsection{Loss Function Design}
The designed loss function of the proposed method includes physical knowledge based forward dynamic loss $L_{m}$, pre-selected inverse dynamics based physiological loss $L_{p}$, boundary loss $L_{b}$, and the force fit loss $L_{f}$, which can be represented as
\begin{equation}
L_{total} = L_m + L_f + \omega (L_p + L_b).
\label{ltotal}
\end{equation}
$\omega$ is the regularization parameter that balances the four terms of the loss function.
\subsubsection{Physical knowledge based forward dynamic loss} $L_{m}$ reflects underlying relationships among the muscle activation and joint kinematics in human motion, given by
\begin{equation}
\begin{aligned}
L_{m} = \frac{1}{T} \sum_{t=1}^{T} (M(q_t)\ddot{q_t} + C(q_t, \dot{q_t}) + G(q_t) - \tau_t)^{2},
\label{r1}
\end{aligned}
\end{equation}
where $M(q_t), C(q_t, \dot{q_t})$ and $G(q_t)$ are the mass matrix, the Centrifugal and Coriolis force, and the gravity. 
$\dot{q_t}$ and $\ddot{q_t}$, the joint angular velocity and joint angular acceleration respectively, are inputs to the neural network, derived from the discrete differentiation of joint motion $q_t$, at the time step $t$. Here, $T$ denotes the movement duration.
$\tau_t$ represents the joint torque, which is calculated by the summation of the product of the moment arm and muscle-tendon force:
\begin{equation}
\tau_t =\sum_{n=1}^{N} F_{t,n}^{t}r_{t,n}
\label{torque}
\end{equation}
where $N$ is the number of muscles involved. $r_{t,n}$ is the moment arm of the $n$th muscle. $F_{t,n}^{t} $ is the estimated muscle-tendon force of the $n$th muscle, which is calculated by the Hill muscle model:
\begin{equation}
\begin{aligned}
F^{t}_{t,n} &= (F_{t,n}^{CE} + F_{t,n}^{PE})\cos{\varphi_{t,n}}\\
&= F_{o,n}^{m}(\hat{a}_{t, n}f_{v}(\overline{v}_{t,n})f_{a}(\overline{l}^{m}_{t,n})\\
&+ f_{p}(\overline{l}^{m}_{t,n}))cos{\varphi_{t,n}},
\label{Fmt}
\end{aligned}
\end{equation}
where $F_{t,n}^{CE}$ is the active force generated by the muscle contraction and $F_{t,n}^{PE}$ represents the passive force generated by the muscle stretch. Muscle activation ($\hat{a}_{t,n}$) is the output of the neural network.
$f_{a}(\overline{l}^{m}_{t,n})$, $f_{v}(\overline{v}_{t,n})$ and $f_{p}(\overline{l}^{m}_{t,n})$ interpret the force-length-velocity characteristics relating to $\hat{a}_{t, n}$, normalized muscle length $\overline{l}^{m}_{t,n}$ and normalized contraction velocity $\overline{v}_{t,n}$. The isometric muscle force is denoted as $F_{o,n}^{m}$. 
The pennation angle $\varphi_{t,n}$ is the angle between the orientation of the muscle fiber and tendon, and the pennation angle at the current muscle fiber length $l^{m}_{t,n}$ can be calculated by
\begin{equation}
\varphi_{t,n} = \sin^{-1}(\frac{l^m_{o,n}\sin{\varphi_{o,n}}}{l^{m}_{t,n}})
\label{fi}
\end{equation} 
where $l^m_{o,n}$ represents the optimal muscle fiber length, and $\varphi_{o,n}$ denotes the optimal pennation angle. The muscle length $l^m_{t,n}$ can be updated by
\begin{equation}
l^m_{t,n} = (l^{mt}_{t,n} - l^{t}_{t,n}){\cos^{-1}{\varphi_{t,n}}}.
\label{lm}
\end{equation}
The total muscle-tendon length is denoted as $l^{mt}_{t,n}$, the tendon length as $l^{t}_{t,n}$.

Before the model training begins, all the physiological parameters, including the previously mentioned $F_{o,n}^{m}, {l}_{o,n}^{m}, \varphi_{o,n}$, as well as the tendon slack length 
$l^{t}_{s, n}$ and the maximum contraction velocity $v_{o,n}$, need to be initialized by linear scaling based on the initial values of the generic model from OpenSim. 
The muscle-tendon length $l^{mt}_{t,n}$ and moment arm $r_{t,n}$ are approximated using a higher-order polynomial function against $q_t$, exported from OpenSim \cite{10.1371/journal.pcbi.1006223}.

\subsubsection{Pre-selected inverse dynamics based physiological loss} 
Based on the MSK inverse dynamics approach, $L_{p}$ is formulated from performance criteria used in the optimization process. This loss function guides the training to refine predicted muscle activation patterns, ensuring they correspond with observed human movements and remain physiologically feasible, even without labeled data.
This study designs $L_p$ to minimize the $\beta$th power of muscle activation.
This formulation can be expressed as follows
\begin{equation}
L_{p}= \frac{1}{T} \sum_{t=1}^{T}\sum_{n=1}^{N} (\hat{a}_{t, n})^\beta 
\end{equation}
where $\hat{a}_{t, n}$ represents the muscle activation of the $n$th muscle at time step $t$. 
Here, the parameter $\beta$ is a constant that remains fixed during training, and we set it as $2$ in this paper~\cite{ACKERMANN20101055, 10.3389/fbioe.2022.1002731, ZARGHAM2019223}.
Similar to inverse dynamics approaches, at each sampling point, our objective is to minimize muscle activation levels while maintaining congruence with observed movements. 

\subsubsection{Boundary Loss}\label{Boundary}
$L_{b}$ is designed to ensure that the predictions of muscle activations remain within a practical and reasonable range, from $0.01$ to $1$. 
Boundary loss functions typically operate by penalizing predictions that exceed predetermined limits. The individual loss function can be defined as handling the loss for a single prediction relative to boundary conditions. 
This is represented by an auxiliary function named $\gamma$, which computes the loss for each prediction when boundary conditions are violated
\begin{equation}
\gamma(x) = \left(\max(0, 0.01 - x)\right)^2 + \left(\max(0, x - 1)\right)^2.
\label{boundary-1}
\end{equation}
The overall boundary loss function across all time steps and muscles can be expressed as
\begin{equation}
L_{b} = \frac{1}{T} \sum_{t=1}^{T} \sum_{n=1}^{N}\gamma(\hat{a}_{t, n}).
\label{boundary-2}
\end{equation}
This loss term ensures that the predicted muscle activation remains within biologically realistic bounds throughout the training process, thereby preventing unreliable predictions.

\subsubsection{Force Fit Loss} There is also an implicit relationship between the predicted muscle forces $\hat{F}_{t,n}^t$ and the muscle-tendon force $F_{t,n}^{t}$ calculated by the embedded MSK dynamics model. 
Thus, $L_{f}$ is designed for estimating muscle forces by minimizing the difference between $\hat{F}_{t,n}^t$ and $F_{t,n}^{t}$, which can be written as
\begin{equation}
L_{f}= \frac{1}{T} \sum_{t=1}^{T}\sum_{n=1}^{N}(\hat{F}_{t,n}^t - F_{t,n}^{t})^2.
\end{equation}

\section{Dataset and Experimental Settings}\label{Dataset}
In this section, data collection and preprocessing are first detailed, followed by baseline methods initialization and evaluation criteria, respectively.
\subsection{Self-Collected Dataset}
The experiment is approved by the Engineering and Physical Sciences Faculty Research Ethics Committee of the University of Leeds (LTELEC-001).
Six subjects participated in this experiment, between the ages of $24$ and $30$. The consent forms are signed by all subjects. 
We collected the subject’s weight and measured the distance from the knee to the ankle to determine the lower leg length, which is crucial for calculating the moment of inertia. 
Subjects participated in a knee bending experiment in a controlled lab setting, using a VICON Motion System at $250$ Hz. 
Reflective markers were placed according to the Plug-in-Gait lower body model to accurately track the movements of the knee joints.
The participants started in a standing position, performed knee flexion to about $90^\circ$, and then returned to the initial position. This bending cycle was repeated four times (Trail $1-4$) per subject.
Each trial lasted approximately $12$ seconds, resulting in around $12,000$ time points in total ($4$ trails).

The marker data were first preprocessed and then analyzed using OpenSim. Specifically, the gait2392 model was scaled to match individual anthropometry using static marker data, with adjusted segment dimensions and muscle properties. Subsequently, joint angles were computed through the Inverse Kinematics tool, followed by the SO tool to estimate muscle activations and forces as experimental ground truths\cite{4352056}.

In the case of knee bending, we focused on five primary muscle-tendon units known to be significantly involved in this motion: Biceps Femoris Short Head (BFS), Biceps Femoris Long Head (BFL), Semitendinosus (SEMT), Medial Gastrocnemius (MG) and Lateral Gastrocnemius (LG). These muscles were selected based on their high activation levels, as indicated by the SO simulation results. 
The physiological parameters of each muscle were obtained from scaled MSK models for the development of integrated loss function.

\subsection{Benchmark Dataset}
This dataset comprises sEMG and the kinematic data from the elbows of ten subjects (six males and four females) performing various exercises, including FE, Pronation-Supination, FE with a 3lb dumbbell, FE with a 5lb dumbbell, etc. 
In this study, we exclusively analyzed the FE movements of four male and two female subjects. Data collection involved subjects standing with arms at their sides while performing movements. Joint angles were estimated by the device's integrated IMUs \cite{toro_ossaba_2023_7946782}. 
For each subject, we extracted $10,000$ time points for analysis.
The moment of inertia was calculated using the weight and arm length data provided. Muscle forces and muscle activations, determined by the SO tool from a scaled generic upper limb model, served as ground truths for the experiments (similar procedures as described in the Self-Collected Dataset).

In the case of elbow FE, we focused on five key muscle-tendon units: Triceps brachii long head (TRIl), Triceps brachii medial head (TRIm), Triceps brachii lateral head (TRIlat), Biceps brachii long head (BICl), Biceps brachii short head (BICs). 
The physiological parameters of each muscle were obtained from scaled MSK models for the loss function design as well.

\begin{figure*}[ht]
    \centering
    \includegraphics[width=0.95\textwidth]{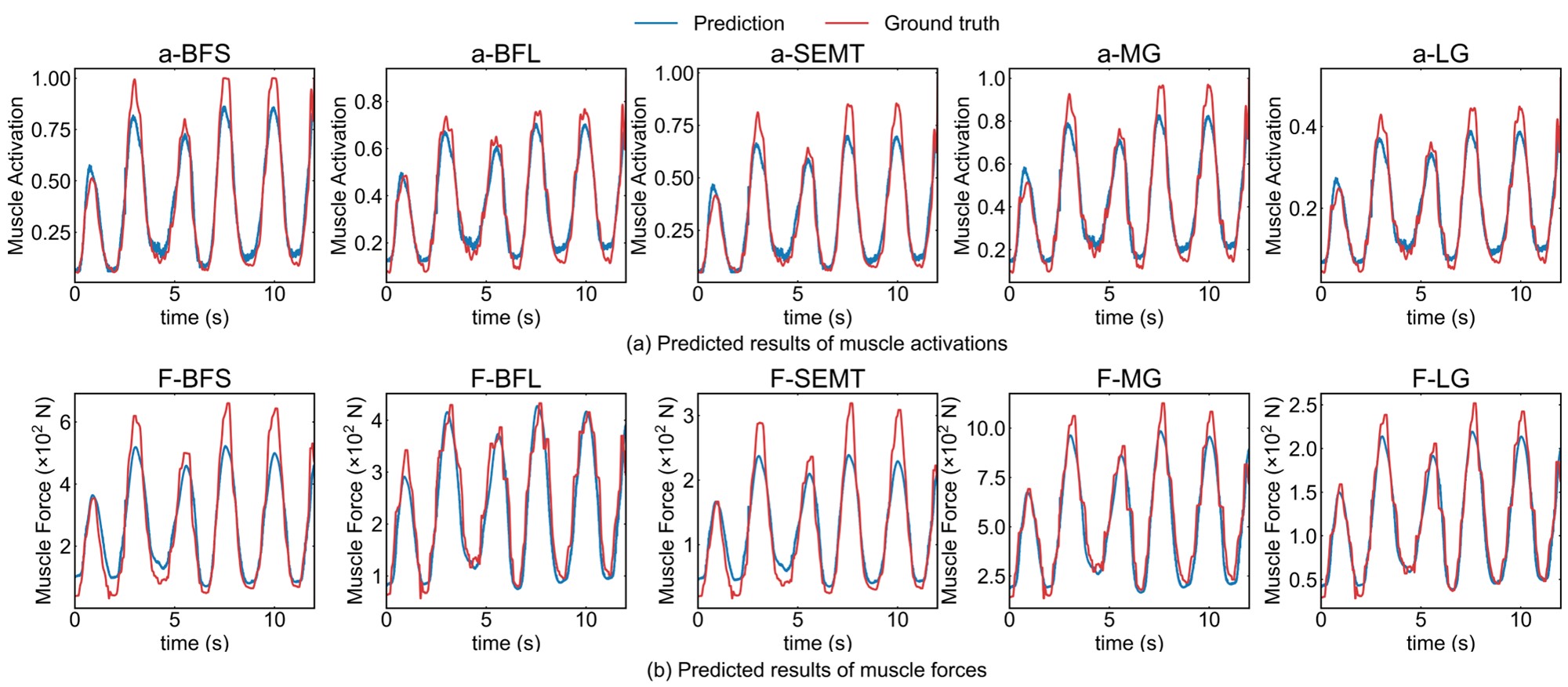}
    \caption{Representative results of the knee bending case through the proposed method. The predicted outputs include the muscle activations and muscle forces of the five main Muscles (BFS, BFL, SEMT, MG, LG).}
    \label{knee-representative}
\end{figure*}

\subsection{Data Preprocessing and Segmentation}

In the data preprocessing phase, the integrated Kalman gap filling tool was used to address the dropout of markers from the raw position data. Joint angles were calculated from the processed marker position data, after which velocity and acceleration were obtained via numerical differentiation.

The preprocessed joint kinematics data were then prepared to input into the BiGRU network using a sliding window technique. Specifically, continuous data were segmented into overlapping blocks by moving a window of $25$ time points with a stride of $2$ time points. Thus, each input sample consists of a $25\times3$ feature matrix, where the three dimensions represent the joint angle, velocity, and acceleration, respectively.

For both two datasets, we adopted an $80-20$ split ratio, where $80\%$ of the data from each subject was used for training and $20\%$ for testing.

\subsection{Baseline Methods and Initialization}
To evaluate the effectiveness of the BiGRU backbone in our proposed method, we compare it with three alternative backbones as the baseline methods in Section IV-A:
Recurrent Neural Network (RNN), Temporal Convolutional Network (TCN), and Long Short-Term Memory (LSTM).
The RNN model consists of two layers, each with $64$ hidden units, and a dropout rate of $0.3$ is applied between layers to enhance generalization.
The TCN model comprises three convolutional layers with channel configurations of $[16, 32, 32]$; each layer uses convolutions with a kernel size of $3$ and dilation factors of $[1, 2, 8]$ to expand the receptive field, effectively capturing temporal features at different time scales. 
A dropout rate of $0.3$ is applied after each convolutional layer to prevent overfitting.
The LSTM model consists of two layers, each with $64$ hidden units, employing sigmoid activation for the gates, tanh for the cell inputs, and a dropout layer of $0.3$ is applied between layers to prevent overfitting.
The RNN, TCN, and LSTM models each incorporate a fully connected block before the output layer to facilitate feature integration and output regression.

To further evaluate the effectiveness of our proposed knowledge-based loss functions, we conducted comparisons with the traditional supervised data-driven method in Section IV-D. The supervised method used muscle activations ($a^o_{t,n}$) and forces ($F_{t,n}^{t,o}$) computed by OpenSim's SO tool as ground truth labels, and employed conventional data-driven mean square error (MSE) loss functions ($MSE(F)$ and $MSE(a)$) to minimize the difference between predictions ($\hat{a}_{t,n}, \hat{F}_{t,n}^{t}$) and these ground truth values.
\begin{equation}
\text{MSE}(F) = \frac{1}{T}\sum_{t=1}^{T}\sum_{n=1}^{N}(F_{t,n}^{t,o} - \hat{F}_{t,n}^{t})^2
\label{msef}
\end{equation}

\begin{equation}
\text{MSE}(a) = \frac{1}{T}\sum_{t=1}^{T}\sum_{n=1}^{N}(a^o_{t,n} - \hat{a}_{t,n})^2
\label{msea}
\end{equation}

To compare with SO-derived ground truths from OpenSim, $\beta$ was set to $2$ for the pre-selected inverse dynamics based physiological loss $L_P$ in all models, including our proposed method.

\subsection{Evaluation Criteria}
Two applied metrics, root mean square error (RMSE) and the coefficient of determination ($R^2$) are used as evaluation metrics to quantify the performance of the regression accuracy, in the experiments. Specifically, the RMSE is defined as:
\begin{equation}
RMSE = \sqrt{\frac{1}{M}\sum_{m=1}^{M}(Y_m - \hat{Y_m})^2}
\end{equation}
where $M$ represents the total number of samples in the evaluation dataset after applying the sliding window segmentation, and $Y_{m}$ and $\hat{Y_{m}}$ denote the actual values and predicted values at $m$th sample, respectively.

The $R^2$ value is calculated using the following formula:
\begin{equation}
\label{r2}
R^2 = 1 - \frac{\sum_{m=1}^{M}(Y_m - \hat{Y_m})^2}{\sum_{m=1}^{M}(Y_m - \overline{Y_m})^2}
\end{equation}
where $\overline{Y_m}$ indicates the mean value of all the samples in the evaluation dataset.  
An $R^2$ value close to $1$ indicates a perfect estimation, while a negative $R^2$ suggests that the sum of the squared estimation errors exceeds the variance of the ground truth.

\begin{figure*}
    \centering
    \includegraphics[width=0.95\textwidth]{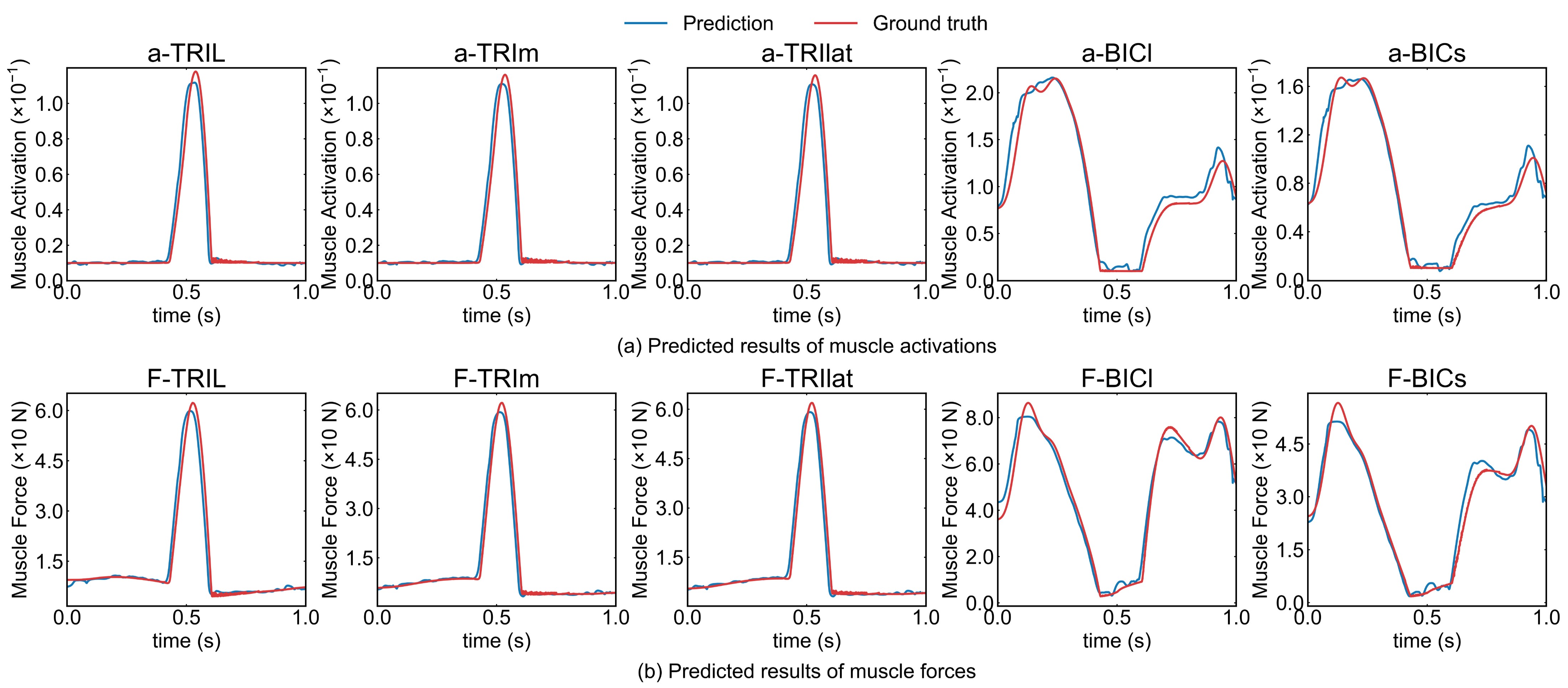}
    \caption{Representative results of the elbow FE case through the proposed method. The predicted outputs include the muscle activations and muscle forces of the five main Muscles (TRIl, TRIm, TRIlat, BICl, BICs).}
    \label{elbow-representative}
\end{figure*}
\begin{figure*}
    \centering
    \includegraphics[width=0.95\textwidth]{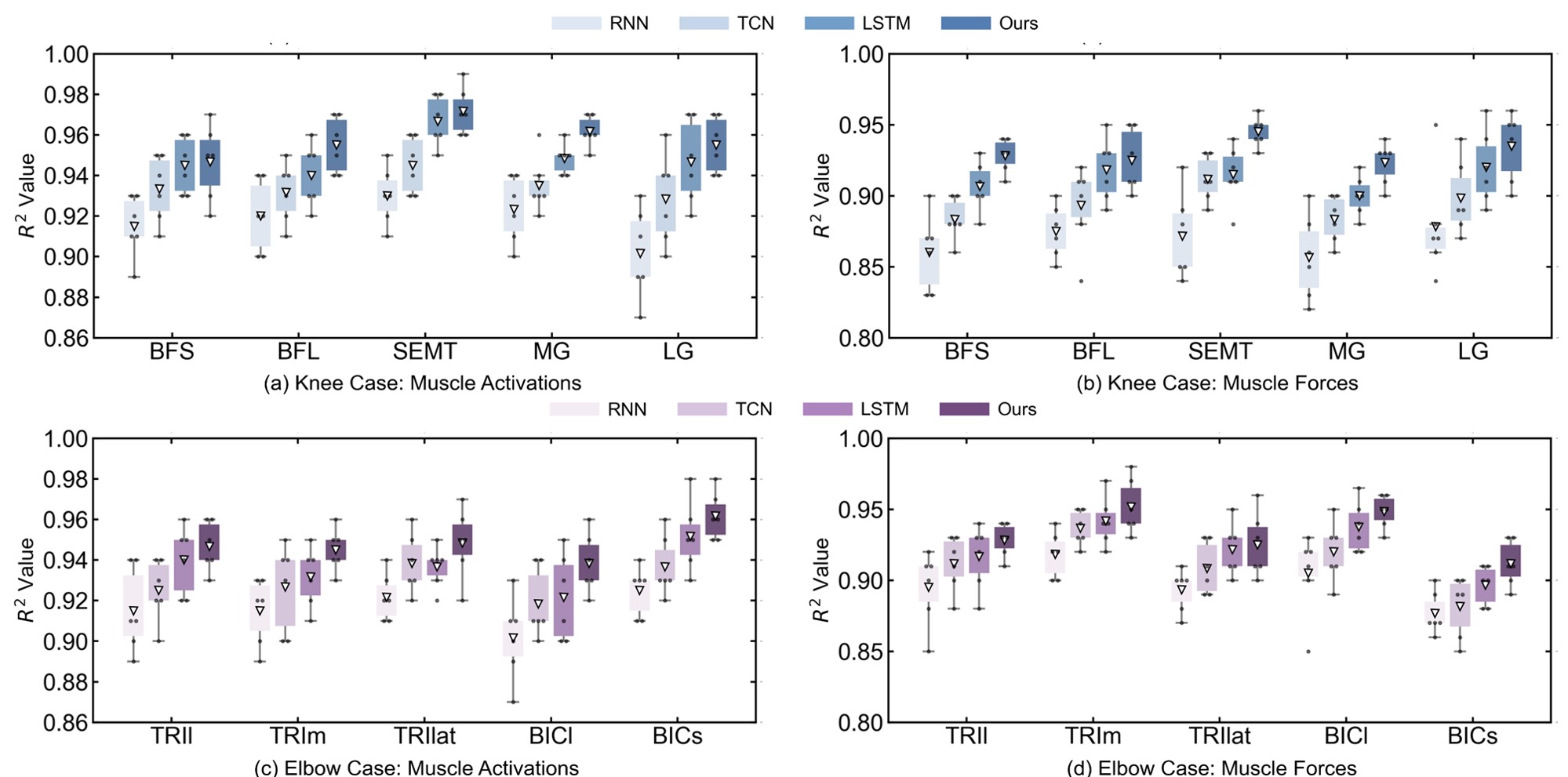}
    \caption{Comparative analysis of average $R^2$ values across subjects using different neural network architectures for two datasets (Knee and Elbow). Results are presented for knee dataset (a-b) and elbow dataset (c-d), showing muscle activation prediction (a,c) and force prediction (b,d). "$\triangledown$" denotes the mean $R^2$ across six subjects, while "\scalebox{0.8}{$\bullet$}" represents individual $R^2$ values for each subject (rounded to two decimal places). }
    \label{r2-combined}
\end{figure*}

\section{Results}\label{results}
In this section, the performance of our proposed method is evaluated using both a self-collected dataset and a benchmark dataset.  
The experimental results are organized into four main subsections. Subsection $A$ presents a comprehensive comparison between our proposed method and the baseline methods.
Subsection $B$ contains an ablation study that examines the individual contributions of each component of the loss function to the training process. Subsection $C$ investigates the optimization of the loss weights to determine their optimal configuration. Finally, Subsection $D$ provides a comparative analysis between our method and the traditional supervised data-driven method. 
Notably, the comparative analysis in Subsection $A$ maintains the identical knowledge-based loss function while varying the architectures of the neural network between the baseline methods to evaluate the effectiveness of the BiGRU backbone in our method.
However, in Subsection $D$, the effectiveness of the knowledge-based loss function is further validated through comparative evaluation of both methods using identical network architectures but trained with distinct loss functions.
All experiments are implemented using the PyTorch framework and executed on a laptop equipped with a GeForce RTX $3070$Ti graphics card and $16$ GB of RAM.

\subsection{Overall Comparison and Architectural Sensitivity Analysis}
\begin{figure*}
    \centering
    \includegraphics[width=0.95\textwidth]{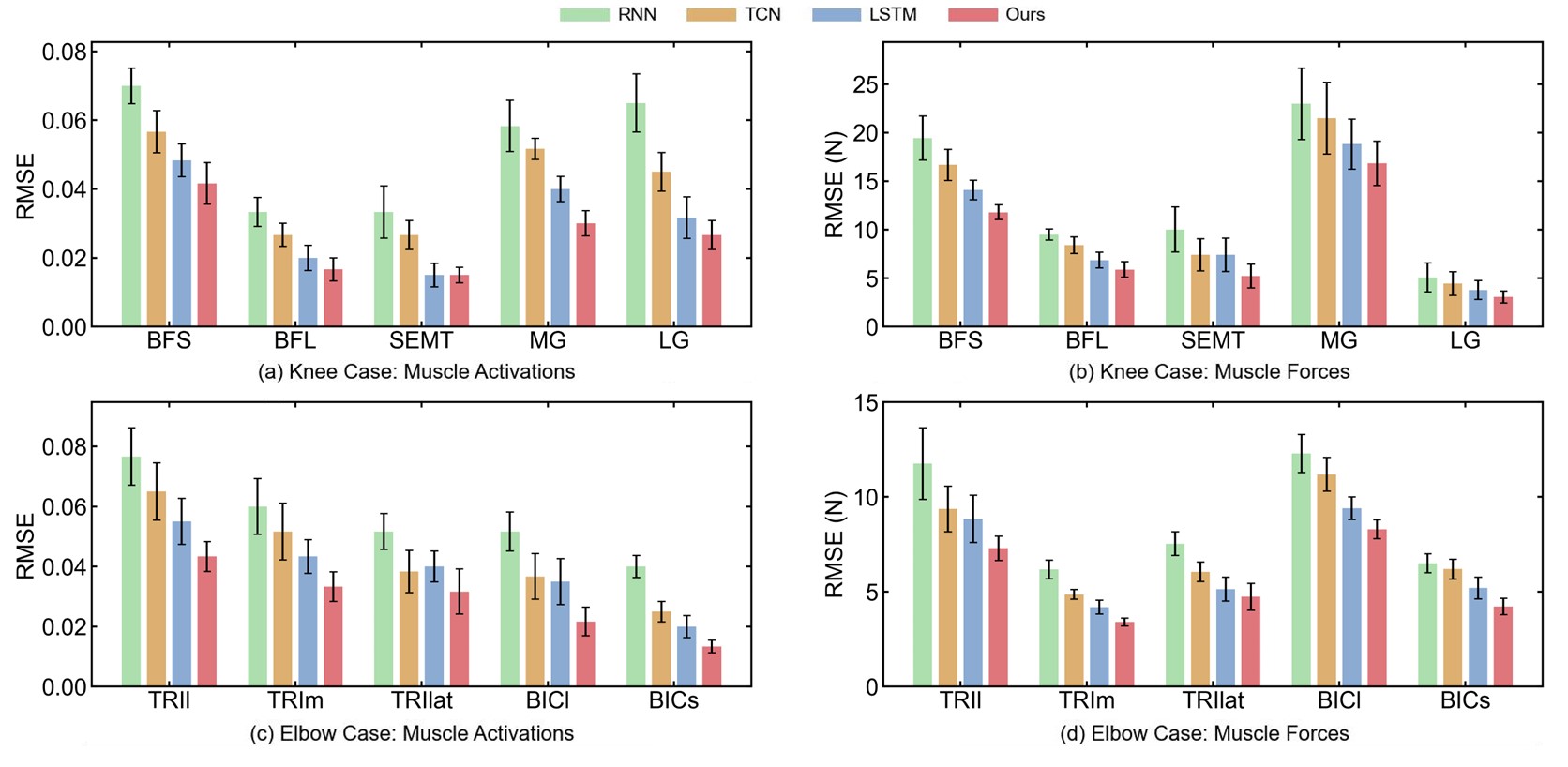}
    \caption{Comparative analysis of average RMSEs across subjects using different neural network architectures for two datasets (Knee and Elbow). Results from knee dataset (a-b) and elbow dataset (c-d) show muscle activation prediction (a,c) and force prediction (b,d).}
    \label{rmses-combined}
\end{figure*}

Figs. \ref{knee-representative} and \ref{elbow-representative} show representative results of our method for predicting muscle activations and forces in knee bending (BFS, BFL, SEMT, MG, LG) and elbow FE (TRIl, TRIm, TRIlat, BICl, BICs) cases, respectively. According to these two figures, the proposed method could predict muscle activations and muscle forces well in both datasets.

To better demonstrate the superiority of our method's neural network architecture, an overall comparative analysis of various backbone networks, including our model and other baseline models, was conducted across two datasets.
Fig. \ref{r2-combined}(a,b) illustrates the $R^2$ comparison using box plots with error bars in the knee bending case. The boxes represent the distribution of $R^2$ values across six subjects and the $\triangledown$ denotes the mean $R^2$ values. The results are presented for five muscles in terms of both activation prediction (a-BFS to a-LG) and force prediction (F-BFS to F-LG), with higher values indicating superior prediction accuracy.
Fig. \ref{rmses-combined}(a,b) visualizes the six-subject averaged RMSE values with error bars, where lower values indicate better performance in the knee bending case.
Specifically, RNN demonstrated its overall $R^2$ values and RMSEs were generally inferior to the other baseline models, suggesting its limitation in handling complex long-term temporal dependencies.
While TCN showed improved performance over RNN due to its effective local temporal pattern extraction, it generally fell short of our model and LSTM. However, TCN demonstrated comparable performance in specific cases (F-SEMT).
LSTM emerged as the strongest baseline method, achieving accuracy comparable to our method (e.g., a-SEMT, F-BFL) while maintaining consistent stability across most cases.
Our method consistently outperformed all baselines, achieving the highest $R^2$ values in both activation and force predictions with the smallest variations, compared to baseline methods in the knee bending case.

In the elbow FE case, it is observed that our model consistently achieves the highest $R^2$ and lowest RMSE values from Fig. \ref{r2-combined}(c,d) and Fig. \ref{rmses-combined}(c,d), as well.
RNN again showed the weakest performance among the baselines. Although TCN's performance was closer to LSTM compared to the knee bending case, LSTM still exhibited better overall results. However, even as the best-performing baseline, LSTM showed notably lower accuracy than our method.

Overall, our BiGRU-based model outperformed the baseline models and consistently achieved accurate predictions across all outputs in both datasets. 
These comprehensive experiments validate both the effectiveness of the BiGRU architecture and demonstrate the adaptability and generalizability of the proposed framework.

\subsection{Analysis of Loss Components}
\begin{table}
\caption{Ablation of our proposed method (knee case)}
\footnotesize
\begin{tabularx}{0.48\textwidth}{@{}>{\centering\arraybackslash}X>{\centering\arraybackslash}X>{\centering\arraybackslash}X>{\centering\arraybackslash}X>{\centering\arraybackslash}X>{\centering\arraybackslash}X@{}}
\cline{1-6}
$L_m$\rule{0pt}{3ex} & $L_f$ & $L_p$ & $L_b$ & $R^2$(a) & $R^2$(F) \\ \cline{1-6}
$\surd$ & $\surd$ & $\surd$ & &0.88 & 0.81 \\
$\surd$ & $\surd$ &  & $\surd$&0.45 & 0.19 \\
$\surd$ &  &$\surd$  & $\surd$&0.93 &-3.71 \\
& $\surd$ &$\surd$ & $\surd$ &0.15 &-0.37 \\
$\surd$ & $\surd$ & $\surd$ & $\surd$ &0.96 &0.93 \\ \cline{1-6}
\end{tabularx}
\label{Ablations-table}
\end{table}

\begin{figure*}
\centering
    \includegraphics[width=0.95\textwidth]{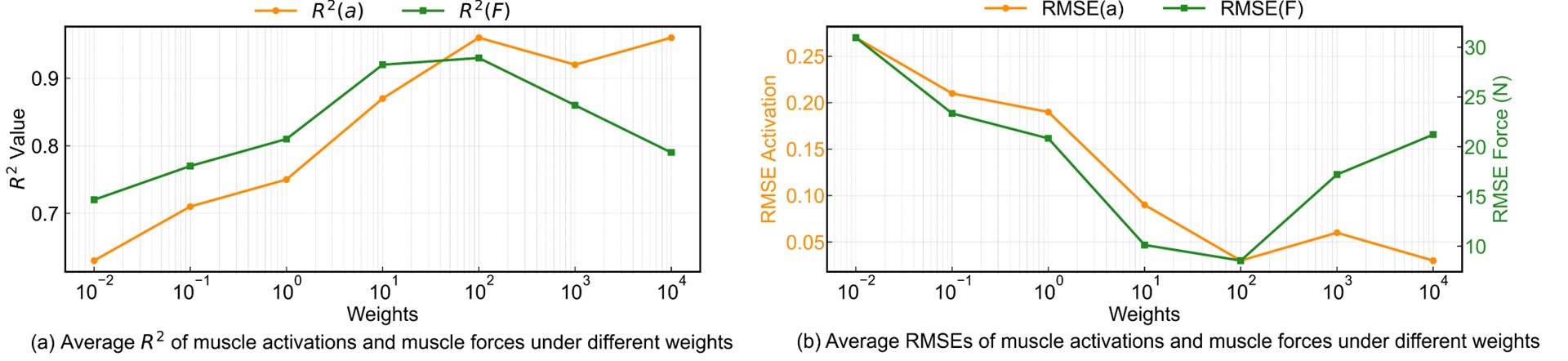}
    \caption{Average $R^2$ and RMSE values of muscle activations and muscle forces across five muscles in knee bending case under different weights}
    \label{weights}
\end{figure*}

To thoroughly understand the individual contribution of each loss term to model performance, we conducted a detailed ablation study on the knee bending dataset as a representative example. 
The experiments were performed by systematically excluding each component of the loss function and retraining the model. The model's accuracy is evaluated using $R^2(a)$ and $R^2(F)$, representing the average $R^2$ values for muscle activations and forces across all five muscles (BFS, BFL, SEMT, MG, LG), respectively. The experimental results are detailed in Table \ref{Ablations-table}.

According to the table, the removal of $L_m$ leads to a substantial drop in performance ($R^2(a)=0.15$ and $R^2(F)=-0.37$),  a decline significantly sharper than that caused by removing other loss terms. This underscores $L_m$'s essential contribution to the model's prediction accuracy.
Excluding $L_f$ primarily affects $R^2(F)$, which decreases to $-3.71$, highlighting its importance in muscle force prediction, while its impact on muscle activation is comparatively minor.
When $L_b$ is omitted, both metrics show modest decreases ($R^2(a)=0.85$ and $R^2(F)=0.71$), indicating that the model remains effective without $L_b$, although not optimally.
The absence of $L_p$ leads to a sharp decline in prediction accuracy ($R^2(a)=0.45$ and $R^2(F)=0.19$), highlighting the significant impact of $L_p$ on the prediction of both muscle activations and muscle forces. 
Particularly, compared to $L_b, L_f$, the influence of $L_p$ on muscle activations is relatively more substantial.
With all loss components included, the model achieves optimal performance ($R^2(a)=0.96$ and $R^2(F)=0.93$), indicating that the combined effect of these losses optimizes both muscle activation and force predictions.

\subsection{Optimization of Loss Function Weights}
A systematic approach combining grid search and validation techniques was used to determine the optimal value of $\omega$, the crucial parameter for balancing the magnitudes of different loss terms. Our method was evaluated on the knee bending dataset, maintaining the same network architecture and hyperparameters throughout the experiments.
The grid search spanned a range from 1 to 201, increasing in increments of 20, to extensively explore the influence of $\omega$ on model training and validation performance.
For muscle activation prediction assessment, we use $R^2(a)$ and $RMSE(a)$, which represent the average $R^2$ and RMSE values across all included muscles. Similarly, $R^2(F)$ and $RMSE(F)$ are employed to evaluate muscle force estimation, representing the average metrics across all muscles.

In Fig. \ref{weights}(a), both $R^2(a)$ and $R^2(F)$ exhibit an increasing trend as the weight parameter escalates from $10^{-2}$ to $10^{2}$. Beyond $10^{2}$, while $R^2(a)$ experiences a slight increase followed by a subsequent decline, $R^2(F)$ reaches a peak at $10^{2}$ and then declines.
Fig. \ref{weights}(b) illustrates a consistent decrease in $RMSE(a)$ as the weight parameter increases, achieving the lowest value at $10^{2}$. Conversely, $RMSE(F)$ follows a similar trend but with a pronounced increase after its lowest point at $10^{2}$.
Based on the above observations, the optimal value of $\omega$ is determined to be $10^{2}$. At this level, both $RMSE(a)$ and $RMSE(F)$ achieve nearly the lowest values within the search range, while $R^2(a)$ and $R^2(F)$ reach nearly the highest values. Therefore, it can be concluded that within our explored range, $\omega = 10^{2}$ offers the best overall performance of our model.
Future work may explore dynamic adjustment strategies for $\omega$ during training to further enhance the robustness and efficiency of the model.

\subsection{Evaluating the Performance Against The Traditional Supervised Data-driven Method}
\begin{figure}
\centering
    \includegraphics[width=0.45\textwidth]{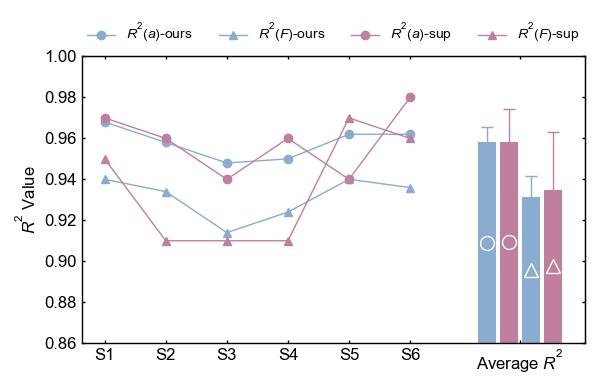}
    \caption{Comparative Analysis of Prediction Performance between Supervised Data-driven and Our Proposed Methods in the Knee Case. Left: subject-wise $R^2$ trajectories (S1-S6); Right: averaged $R^2$ values}
    \label{knee-sup}
\end{figure}

To validate the performance difference between the proposed method and the traditional supervised data-driven method and further evaluate the effectiveness of our proposed knowledge-based loss function, we conducted comprehensive comparisons using identical BiGRU architecture trained with different loss functions: knowledge-based $L_{total}$ (Eq. (\ref{ltotal})) for our method and conventional MSE loss functions (Eq. (\ref{msef}), Eq. (\ref{msea})) for the supervised method. The same data preprocessing and splitting strategy was used as in the overall comparison (Section IV-A).

\begin{figure}
\centering
    \includegraphics[width=0.45\textwidth]{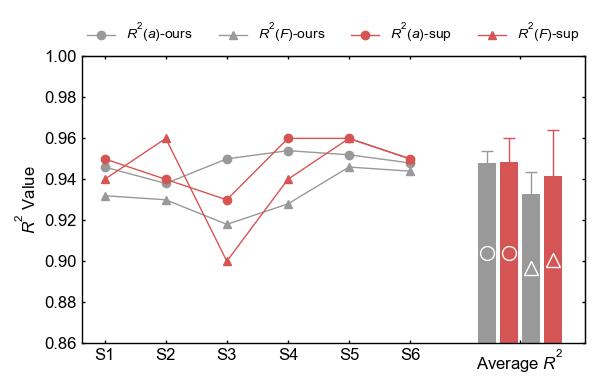}
    \caption{Comparative Analysis of Prediction Performance between Supervised Data-driven and Our Proposed Methods in the ELbow Case. Left: subject-wise $R^2$ trajectories (S1-S6); Right: averaged $R^2$ values}
    \label{elbow-sup}
\end{figure}

Figs. \ref{knee-sup} and \ref{elbow-sup} present comprehensive comparative analyses between our knowledge-based method and the supervised method for both knee bending and elbow FE cases, where $R^2(a)$ and $R^2(F)$ represent the average $R^2$ values across all muscles for muscle activations and forces predictions, respectively.
In the knee bending case (Fig. \ref{knee-sup}), our method demonstrates robust performance with $R^2$ values consistently exceeding 0.90 across all subjects. Notably, for muscle activation prediction, our method achieves comparable or even superior performance compared to the supervised method in S3. The averaged results show that our method maintains highly competitive accuracy ($R^2(a) \approx 0.96$) with the supervised method, while exhibiting smaller standard deviations, indicating enhanced prediction stability.
For the elbow FE case (Fig. \ref{elbow-sup}), a similar pattern of robust performance is observed. The subject-wise trajectories show that our method maintains stable performance across all subjects, with $R^2$ values consistently above $0.92$, while the supervised method shows noticeable fluctuation in $R^2(F)$. The averaged results demonstrate that our method achieves comparable accuracy in activation prediction ($R^2(a) \approx 0.95$) with marginally lower but stable force prediction performance ($R^2(F) \approx 0.93$).

The experimental results demonstrate that our knowledge-based method achieves comparable prediction accuracy to the supervised method while exhibiting enhanced stability across subjects, despite the occasional higher peak performance of the supervised method in individual cases. These findings not only validate the effectiveness of our proposed knowledge-based loss functions in guiding network training but also demonstrate that domain knowledge can effectively substitute for ground truth data.

\section{Discussion}\label{discussions}
This paper introduces a novel knowledge-based deep learning method specifically designed for estimating muscle activations and forces.
In the overall comparison with the baseline methods, our method demonstrates robust performance across two datasets without requiring labeled training data.
In comparison with the traditional supervised data-driven method, our method demonstrates comparable performance and superior stability across both datasets.
Despite these promising results, several limitations exist. Subsequently, we discuss these limitations and explore potential approaches to further enhance the method's performance from multiple perspectives.

Our method is developed based on a known MSK system model to predict muscle forces and muscle activation. 
This approach involves initial modeling of the entire MSK system, capturing the specific dynamics of movement and muscle activation for each muscle-tendon unit, which may vary according to experimental conditions and participant characteristics. 
These crucial elements are integrated into our loss function to enhance predictive accuracy.
Currently, our method is suited for movements involving a single degree of freedom and joint, because of their simpler and more uniform structures which facilitate model transfer and application. 
However, in systems with complex structures and varied movement patterns, this approach requires extensive preliminary modeling and parameter tuning, limiting its broader applicability and flexibility.
In response to these challenges, our future work will aim to enhance our method to better accommodate and analyze complex movements. We plan to develop more versatile modeling techniques and explore algorithms for adaptive learning of dynamical parameters, striving to extend the applicability and improve the predictive accuracy of our model more comprehensively.

Our method is inspired by traditional optimization algorithms from MSK inverse dynamics, to estimate muscle activation and muscle forces. 
This is crucial for analyzing complex movements or when EMG data from specific muscle-tendon units are unobtainable. 
However, like these traditional methods, our approach sometimes lacks a physiological basis, leading to discrepancies between estimated and actual measurements, as evidenced in \cite{TRINLER2018266}.
To enhance our method's applicability and align outputs more closely with physiological realities, we propose integrating our model with measured EMG data. This integration of forward and inverse dynamics, aims to calibrate our model’s outputs, not only ensures the accuracy of estimations but also accommodates the neural control strategies of the subjects. 

Additionally, it is worth noting that our model is designed with flexibility, allowing us to modify the embedded MSK model's structure and parameters according to different research conditions and objectives, including but not limited to the adjustment of the motion equations and the performance criteria.

\section{Conclusion}\label{conclusion}
This paper presents a novel, time-efficient, knowledge-based deep learning method for the direct estimation of muscle activations and forces from joint kinematic data. A BiGRU-based model is utilized as the neural network surrogate, integrating prior physical knowledge from forward dynamics and pre-selected inverse dynamics-based physiological criteria into the loss function to guide training without the need for labeled data. Extensive experiments demonstrate the effectiveness and potential of the proposed method.

\bibliographystyle{ieeetr}
\bibliography{reference}

\begin{thebibliography}{10}

\bibitem{9923768}
Y.~Zhao, K.~Qian, S.~Bo, Z.~Zhang, Z.~Li, G.-Q. Li, A.~A. Dehghani-Sanij, and S.~Q. Xie, ``Adaptive cooperative control strategy for a wrist exoskeleton using model-based joint impedance estimation,'' {\em IEEE/ASME Transactions on Mechatronics}, vol.~28, no.~2, pp.~748--757, 2023.

\bibitem{8627981}
M.~Sartori, J.~van~de Riet, and D.~Farina, ``Estimation of phantom arm mechanics about four degrees of freedom after targeted muscle reinnervation,'' {\em IEEE Transactions on Medical Robotics and Bionics}, vol.~1, no.~1, pp.~58--64, 2019.

\bibitem{9116986}
X.~Jiang, H.~Ren, K.~Xu, X.~Ye, C.~Dai, E.~A. Clancy, Y.-T. Zhang, and W.~Chen, ``Quantifying spatial activation patterns of motor units in finger extensor muscles,'' {\em IEEE Journal of Biomedical and Health Informatics}, vol.~25, no.~3, pp.~647--655, 2021.

\bibitem{10388411}
L.~Zhang, T.~Van~Wouwe, S.~Yan, and R.~Wang, ``Emg-constrained and ultrasound-informed muscle-tendon parameter estimation in post-stroke hemiparesis,'' {\em IEEE Transactions on Biomedical Engineering}, vol.~71, no.~6, pp.~1798--1809, 2024.

\bibitem{9546647}
R.~Zaman, Y.~Xiang, R.~Rakshit, and J.~Yang, ``Hybrid predictive model for lifting by integrating skeletal motion prediction with an opensim musculoskeletal model,'' {\em IEEE Transactions on Biomedical Engineering}, vol.~69, no.~3, pp.~1111--1122, 2022.

\bibitem{10050735}
J.~Tong, A.~V. Subramani, V.~Kote, M.~Baggaley, W.~B. Edwards, and J.~Reifman, ``Effects of stature and load carriage on the running biomechanics of healthy men,'' {\em IEEE Transactions on Biomedical Engineering}, vol.~70, no.~8, pp.~2445--2453, 2023.

\bibitem{10538305}
Z.~Xia, B.~M. Cornish, D.~Devaprakash, R.~S. Barrett, D.~G. Lloyd, A.~H. Hams, and C.~Pizzolato, ``Prediction of achilles tendon force during common motor tasks from markerless video,'' {\em IEEE Transactions on Neural Systems and Rehabilitation Engineering}, vol.~32, pp.~2070--2077, 2024.

\bibitem{9756369}
R.~Hu, X.~Chen, H.~Zhang, X.~Zhang, and X.~Chen, ``A novel myoelectric control scheme supporting synchronous gesture recognition and muscle force estimation,'' {\em IEEE Transactions on Neural Systems and Rehabilitation Engineering}, vol.~30, pp.~1127--1137, 2022.

\bibitem{10366325}
I.~Loi, E.~I. Zacharaki, and K.~Moustakas, ``Multi-action knee contact force prediction by domain adaptation,'' {\em IEEE Transactions on Neural Systems and Rehabilitation Engineering}, vol.~32, pp.~122--132, 2024.

\bibitem{NeuromusculoskeletalModel}
T.~S. Buchanan, D.~G. Lloyd, K.~Manal, and T.~F. Besier, ``Neuromusculoskeletal modeling: Estimation of muscle forces and joint moments and movements from measurements of neural command,'' {\em Journal of Applied Biomechanics}, vol.~20, no.~4, pp.~367 -- 395, 2004.

\bibitem{9258965}
Y.~Zhao, Z.~Zhang, Z.~Li, Z.~Yang, A.~A. Dehghani-Sanij, and S.~Xie, ``An {EMG}-driven musculoskeletal model for estimating continuous wrist motion,'' {\em IEEE Transactions on Neural Systems and Rehabilitation Engineering}, vol.~28, no.~12, pp.~3113--3120, 2020.

\bibitem{8681204}
A.~Zonnino and F.~Sergi, ``Model-based estimation of individual muscle force based on measurements of muscle activity in forearm muscles during isometric tasks,'' {\em IEEE Transactions on Biomedical Engineering}, vol.~67, no.~1, pp.~134--145, 2020.

\bibitem{10.1093/ptj/61.1.7}
J.~Perry, C.~S. Easterday, and D.~J. Antonelli, ``{Surface Versus Intramuscular Electrodes for Electromyography of Superficial and Deep Muscles},'' {\em Physical Therapy}, vol.~61, pp.~7--15, 01 1981.

\bibitem{9672677}
K.~J. Bennett, C.~Pizzolato, S.~Martelli, J.~S. Bahl, A.~Sivakumar, G.~J. Atkins, L.~B. Solomon, and D.~Thewlis, ``Emg-informed neuromusculoskeletal models accurately predict knee loading measured using instrumented implants,'' {\em IEEE Transactions on Biomedical Engineering}, vol.~69, no.~7, pp.~2268--2275, 2022.

\bibitem{10197487}
K.~Ayusawa, A.~Murai, R.~Sagawa, and E.~Yoshida, ``Fast inverse kinematics based on pseudo-forward dynamics computation: Application to musculoskeletal inverse kinematics,'' {\em IEEE Robotics and Automation Letters}, vol.~8, no.~9, pp.~5775--5782, 2023.

\bibitem{SEIREG197589}
A.~Seireg and R.~Arvikar, ``The prediction of muscular load sharing and joint forces in the lower extremities during walking,'' {\em Journal of Biomechanics}, vol.~8, no.~2, pp.~89--102, 1975.

\bibitem{CROWNINSHIELD1981793}
R.~D. Crowninshield and R.~A. Brand, ``A physiologically based criterion of muscle force prediction in locomotion,'' {\em Journal of Biomechanics}, vol.~14, no.~11, pp.~793--801, 1981.

\bibitem{HEINTZ2007279}
S.~Heintz and E.~M. Gutierrez-Farewik, ``Static optimization of muscle forces during gait in comparison to emg-to-force processing approach,'' {\em Gait \& Posture}, vol.~26, no.~2, pp.~279--288, 2007.

\bibitem{ZARGHAM2019223}
A.~Zargham, M.~Afschrift, J.~{De Schutter}, I.~Jonkers, and F.~{De Groote}, ``Inverse dynamic estimates of muscle recruitment and joint contact forces are more realistic when minimizing muscle activity rather than metabolic energy or contact forces,'' {\em Gait \& Posture}, vol.~74, pp.~223--230, 2019.

\bibitem{VEERKAMP2021110530}
K.~Veerkamp, N.~Waterval, T.~Geijtenbeek, C.~Carty, D.~Lloyd, J.~Harlaar, and M.~{van der Krogt}, ``Evaluating cost function criteria in predicting healthy gait,'' {\em Journal of Biomechanics}, vol.~123, p.~110530, 2021.

\bibitem{TRINLER201955}
U.~Trinler, H.~Schwameder, R.~Baker, and N.~Alexander, ``Muscle force estimation in clinical gait analysis using anybody and {OpenSim},'' {\em Journal of Biomechanics}, vol.~86, pp.~55--63, 2019.

\bibitem{9927168}
C.~Lin, X.~Chen, W.~Guo, N.~Jiang, D.~Farina, and J.~Su, ``A bert based method for continuous estimation of cross-subject hand kinematics from surface electromyographic signals,'' {\em IEEE Transactions on Neural Systems and Rehabilitation Engineering}, vol.~31, pp.~87--96, 2023.

\bibitem{10138591}
T.~Bao, C.~Wang, P.~Yang, S.~Q. Xie, Z.-Q. Zhang, and P.~Zhou, ``Lstm-ae for domain shift quantification in cross-day upper-limb motion estimation using surface electromyography,'' {\em IEEE Transactions on Neural Systems and Rehabilitation Engineering}, vol.~31, pp.~2570--2580, 2023.

\bibitem{9970372}
J.~Zhang, Y.~Zhao, F.~Shone, Z.~Li, A.~F. Frangi, S.~Q. Xie, and Z.-Q. Zhang, ``Physics-informed deep learning for musculoskeletal modeling: Predicting muscle forces and joint kinematics from surface emg,'' {\em IEEE Transactions on Neural Systems and Rehabilitation Engineering}, vol.~31, pp.~484--493, 2023.

\bibitem{10398450}
J.~Fan and X.~Hu, ``Towards efficient neural decoder for dexterous finger force predictions,'' {\em IEEE Transactions on Biomedical Engineering}, vol.~71, no.~6, pp.~1831--1840, 2024.

\bibitem{10332235}
H.~Zhang, H.~Qu, L.~Teng, and C.-Y. Tang, ``Lstm-msa: A novel deep learning model with dual-stage attention mechanisms forearm emg-based hand gesture recognition,'' {\em IEEE Transactions on Neural Systems and Rehabilitation Engineering}, vol.~31, pp.~4749--4759, 2023.

\bibitem{10.3389/fncom.2021.759489}
A.~Nasr, K.~A. Inkol, S.~Bell, and J.~McPhee, ``Inversemusclenet: Alternative machine learning solution to static optimization and inverse muscle modeling,'' {\em Frontiers in Computational Neuroscience}, vol.~15, 2021.

\bibitem{LIANG2024112093}
W.~Liang, H.~{Muhammad Rehan Afzal}, Y.~Qiao, A.~Fan, F.~Wang, Y.~Hu, and P.~Yang, ``Estimation of electrical muscle activity during gait using inertial measurement units with convolution attention neural network and small-scale dataset,'' {\em Journal of Biomechanics}, vol.~167, p.~112093, 2024.

\bibitem{RN118}
T.~T. Dao, ``From deep learning to transfer learning for the prediction of skeletal muscle forces,'' {\em Medical \& Biological Engineering \& Computing}, vol.~57, no.~5, pp.~1049--1058, 2019.

\bibitem{RN8}
L.~Rane, Z.~Ding, A.~H. McGregor, and A.~M.~J. Bull, ``Deep learning for musculoskeletal force prediction,'' {\em Annals of Biomedical Engineering}, vol.~47, no.~3, pp.~778--789, 2019.

\bibitem{RN9}
A.~S. Oliveira and C.~I. Pirscoveanu, ``Implications of sample size and acquired number of steps to investigate running biomechanics,'' {\em Scientific Reports}, vol.~11, no.~1, p.~3083, 2021.

\bibitem{10.3389/fbioe.2020.603907}
J.~Holder, U.~Trinler, A.~Meurer, and F.~Stief, ``A systematic review of the associations between inverse dynamics and musculoskeletal modeling to investigate joint loading in a clinical environment,'' {\em Frontiers in Bioengineering and Biotechnology}, vol.~8, 2020.

\bibitem{RN97}
G.~E. Karniadakis, I.~G. Kevrekidis, L.~Lu, P.~Perdikaris, S.~Wang, and L.~Yang, ``Physics-informed machine learning,'' {\em Nature Reviews Physics}, vol.~3, no.~6, pp.~422--440, 2021.

\bibitem{10.1371/journal.pcbi.1006223}
A.~Seth, J.~L. Hicks, T.~K. Uchida, A.~Habib, C.~L. Dembia, J.~J. Dunne, C.~F. Ong, M.~S. DeMers, A.~Rajagopal, M.~Millard, S.~R. Hamner, E.~M. Arnold, J.~R. Yong, S.~K. Lakshmikanth, M.~A. Sherman, J.~P. Ku, and S.~L. Delp, ``{OpenSim}: Simulating musculoskeletal dynamics and neuromuscular control to study human and animal movement,'' {\em PLOS Computational Biology}, vol.~14, no.~7, pp.~1--20, 2018.

\bibitem{ACKERMANN20101055}
M.~Ackermann and A.~J. {van den Bogert}, ``Optimality principles for model-based prediction of human gait,'' {\em Journal of Biomechanics}, vol.~43, no.~6, pp.~1055--1060, 2010.

\bibitem{10.3389/fbioe.2022.1002731}
I.~Luis, M.~Afschrift, F.~De~Groote, and E.~M. Gutierrez-Farewik, ``Evaluation of musculoskeletal models, scaling methods, and performance criteria for estimating muscle excitations and fiber lengths across walking speeds,'' {\em Frontiers in Bioengineering and Biotechnology}, vol.~10, 2022.

\bibitem{4352056}
S.~L. Delp, F.~C. Anderson, A.~S. Arnold, P.~Loan, A.~Habib, C.~T. John, E.~Guendelman, and D.~G. Thelen, ``Opensim: Open-source software to create and analyze dynamic simulations of movement,'' {\em IEEE Transactions on Biomedical Engineering}, vol.~54, no.~11, pp.~1940--1950, 2007.

\bibitem{toro_ossaba_2023_7946782}
A.~Toro-Ossaba and J.~C. Tejada, ``Emg elbow dataset.'' Zenodo, May 2023.

\bibitem{TRINLER2018266}
U.~Trinler, F.~Leboeuf, K.~Hollands, R.~Jones, and R.~Baker, ``Estimation of muscle activation during different walking speeds with two mathematical approaches compared to surface emg,'' {\em Gait \& Posture}, vol.~64, pp.~266--273, 2018.

\end{thebibliography}

\vspace{12pt}
\color{red}

\end{document}